\documentclass[arguments]{aastex631}

\accepted{19th October 2023}

\submitjournal{Apj}

\shorttitle{FFP mass function and parallax event rate}
\shortauthors{Ban}
\graphicspath{{./}{figures/}}
\usepackage{multirow}
\usepackage{bm}

\newcommand{\bes}{Besan\c{c}on}

\newcommand{\jwst}{{\it James Webb Space Telescope}}
\newcommand{\ariel}{{\it Ariel}}
\newcommand{\plato}{{\it PLATO}}
\newcommand{\euclid}{{\it Euclid}}
\newcommand{\nancy}{{\it Roman}}
\newcommand{\kepler}{{\it Kepler}}
\newcommand{\sptz}{{\it Spitzer}}
\newcommand{\moa}{MOA}
\newcommand{\ogle}{OGLE}
\newcommand{\kmt}{KMTNet}
\newcommand{\ts}{\textstyle}
\newcommand{\mlm}{{\it MulensModel}}

\renewcommand\thefootnote{\roman{footnote}}

\begin{document}

\title {The impact of the free-floating planet (FFP) mass function on the event rate for the accurate microlensing parallax determination: application to \euclid{} and \nancy{} parallax observation} 

\author[0000-0002-6079-3335]{Ban, M.}
\affiliation{University of Warsaw Astronomical Observatory, Warsaw, Poland}

\begin{abstract}
A microlensing event is mainly used to search for free-floating planets (FFPs). To estimate the FFP mass and distance via the microlensing effect, a microlensing parallax is one of the key parameters. A short duration of FFP microlensing is difficult to yield a parallax by the observer's motion at a recognisable level, so the FFP microlensing parallax is expected on the simultaneous observation by multiple telescopes. Here, we approach the FFP detection by considering a variation of FFP mass functions and the event rate of accurately measured microlensing parallax. We used our FFP microlensing simulator assuming a parallax observation between upcoming space-based missions (\euclid{} and \nancy{}) with full kinematics. As a result, we confirmed that the event rate of accurately measured microlensing parallax (i.e. within a factor of 2 uncertainty) does not simply follow the number of FFPs at a given mass but the ratio of the FFP population per star. This is because the population ratio determines the optical depth for a given mass and potential sources. In addition, we found that the probability of the event that can estimate the FFP mass and distance within a factor of 2 is not so high: $\sim$40\% of Earth-mass, $\sim$16\% of Neptune-mass, and $\sim$4\% of Jupiter-mass FFP events under our criteria. The probability can be improved by some technical approach such as using high cadence and observation in parallax more than two observers.
\end{abstract}


\section{Introduction} \label{sec:intro}
Gravitational microlensing observation is a method to find isolated planetary-mass objects such as free-floating planets (FFPs). To find the FFP properties via the microlensing effect, it is essential to measure a microlensing parallax ($\pi_E$) which is a value of the stereo-visional observation angle. The accuracy of the microlensing parallax measurement directly affects the accuracy of the lens property estimation. However, the number of FFP event observations is still not enough. In recent decades, microlensing surveys have detected tens of thousands of microlensing events, of which there are only a few published candidates \citep{Mroz2018, Mroz2019a, Mroz2020, Kim2021, Ryu2021, Gould2022}. FFP microlensing events are difficult to observe because of the short event duration and the small event rate. The event rate issue has several views such as the telescope sensitivity limits, FFP population, and the difficulty of the determination whether it is isolated or bound in a wide orbit, and whether it is categorised into planets or low-mass brown dwarfs \citep{Ban2016, Bayo2017, Barclay2017, Delorme2017, Hamolli2013, Hamolli2016, Henderson2016a, Mroz2017, Gould2013a, Gould2013b, Pacucci2013, Penny2017, Sumi2011, Yee2013}. 

The FFP population has been discussed in recent decades. \cite{Sumi2011} first argued that the population of Jupiter-mass FFP is $\sim$1.8 times larger than that of the main-sequence (MS) stars. Later, \cite{Clanton2017} and \cite{Mroz2017} suggested the overestimation of the Jupiter-mass FFP population by calculation and observation, respectively. They provided the population of $\sim$1.4 and $<$0.25 times larger than the MS star population, respectively. Thus, the prediction of the FFP population has been drastically changed. The overall FFP population is still not settled, but some recent research \citep{Gould2022, Johnson2020, Sumi2023} partially agree with each other; especially for Jupiter-mass and Earth-mass FFPs. \cite{Johnson2020} simulated the upcomming space-based mission Nancy Gracy Roman Space Telescope (\nancy{}) and derived the expected FFP population, and \cite{Gould2022} and \cite{Sumi2023} evaluated the latest FFP population from \kmt{} data and \moa{} data, respectively. Aside from the microlensing approach, \cite{Scholz2022} predicted the FFP population from the star-forming clusters observed by \jwst{} that up to 5\% of the population in a cluster could be massive FFPs ($>5M_{Jupiter}$).

%Our previous research showed that the lens mass estimation of FFPs from the microlensing parallax was not accurate enough \citep{Ban2020}. Only $\sim$5\% of the lens mass estimation was plausible with $<$10\% discrepancy. One reason for such low estimation accuracy was that we did not adjust the observer separation to the observers' reference frame. However, according to the event kinematics, it was not the only cause. Besides, \cite{Gould2020} mentioned the probability of two different projected source paths on the microlensing parallax estimation which \cite{Refsdal1966} theoretically introduced. In the case of simultaneous parallax detection by multi-observers, the difference of minimum impact parameters is either of two solutions: $\Delta u_0=u_{0, a} \pm u_{0,b}$ where $a$ and $b$ indicate different telescopes. The probability of taking which solution relies on the event rate and lens population.

Parallax detection helps to estimate the lens properties such as mass and distance \citep{Refsdal1966}. The observer's motion yields a parallax effect, and it appears as the deformation or asymmetry of the light curve.  However, the parallax effect by an observer's motion is likely unrecognizable for FFP microlensing events due to its short duration. \cite{Koshimoto2023} also concluded the difficulty of recognising parallax from the FFP microlensing light curve candidates observed by \moa{}. Hence, simultaneous parallax observation by multi-observers is more effective for FFP microlensing events.  The upcoming space-based missions (\euclid{} and \nancy{}) can yield a simultaneous parallax observation between them and between any ongoing ground-based surveys such as \moa{}, \ogle{}, and \kmt{}.  Some papers, including our previous research, discuss the FFP event configurations including the event rate and the parallax sensitivity \citep{Bachelet2018, Bachelet2022, Ban2016, Ban2020, Hamolli2013, Hamolli2016, Johnson2020, Penny2014, Spergel2013}. Particularly, \cite{Bachelet2019, Bachelet2022} simulated the capability of parallax and mass measurements between \euclid{} and \nancy{} assuming the correspondence of the FFP and bound planet population through detectable microlensing event and two well-known predictions of FFP population: two Dirac delta functions with two peaks at $10M_{\oplus}$ \citep{Mroz2019a} and $2M_{Jupiter}$ \citep{Sumi2011} per main sequence star. In this paper, we assume the parallax observation between \euclid{} and \nancy{} with variations of the FFP mass function referred from \cite{Gould2022, Johnson2020, Sumi2023} and try to approach the relationship between the FFP population, event sensitivity, and accuracy of the microlensing parallax measurement. As a similar research, \cite{Hamolli2017} simulated the FFP microlensing detectability between \kepler{} and \sptz{} by varying the FFP mass function, but the combination of these two telescopes seems to be not suitable for research on the low-mass FFPs because of the wide separation and low observation cadence.

The brief process of our research is as follows. First, we generate the sample artificial events using our microlensing event simulator. Second, \mlm{}, one of the open-source microlensing light curve analysis tools \citep{Poleski2019}, is applied for all sample events to fit the light curve. Third, the microlensing parallax value is found using the result of fitted parameters of the light curve. The event rate of accurately measured microlensing parallax is derived for different FFP mass functions. Hence, the structure of the paper is also following the process. \S\ref{sec:simulator} describes our microlensing event simulator. We summarise the different FFP mass functions used in our research in \S\ref{sec:massfunc} and the generalised light curve fitting to approach the event rate of the accurate estimation of the microlensing parallax in \S\ref{sec:fitting}. The result is in \S\ref{sec:results}, and we analyse the relationship among the estimation accuracy of the microlensing parallax, FFP event rate, and FFP mass functions. In \S\ref{sec:discussion}, we discuss some additional aspects of our results. Finally, in \S\ref{sec:conclusion}, we summarise our results and future research objectives.

%Direct imaging in multi-band likely help to determine whether the lens is a dwarf or an FFP from the chemical compositions of the surrounding discs or atmosphere \citep{Bayo2017, Delorme2017}. However, this approach is generally not applicable to FFPs. Thus, the FFP research relies on microlensing observation. Some upcoming space-based missions will yield more FFP observations \citep{Spergel2013, Penny2014, Ban2016} since the infrared filters, which can cover dimmer sources than the optical filters, are implemented. 

\begin{table}
\centering
\caption{Survey settings of \euclid{} and \nancy{} for our simulator. The \euclid{} configuration is taken from Table 2 of \citet{Penny2013} and \citet{ESA2011}, and the \nancy{} configuration is from \citet{Spergel2013,Spergel2015,Webster2017}. We treat sample source flux in the Johnson-Cousins photometric system throughout our simulation. Therefore, a proxy band (in the parenthesis of the Filter row) is assumed. The threshold amplitude ($A_t$) to make an event alert corresponds to the threshold impact parameter of $u_t \sim 3$ for the point source case.}
\label{tab:survs}
% \begin{tabular}{@{}p{0.7in}p{1.3in}p{0.7in}@{}}
\begin{tabular}{|l|cc|}
\hline
& \euclid{} & \nancy{} \\ \hline
Location & Lagrangian Point 2 & Lagrangian Point 2 \\
Orbital diameter [km] & $\sim0.97\times10^6$ & $\sim1.5\times10^6$ \\
Orbital period [year] & 0.5 & 0.5 \\
Filter & NIPS $H$ ($H$) & W149 ($H$) \\  \hline
$A_{\rm t}$ & 1.02 & 1.02 \\
$m_{\rm sky}$ [mag$/$arcsec$^2$] & 21.5 & 21.5 \\	
$\theta_{\rm psf}$ [arcsec] & 0.4 & 0.4 \\
$m_{\rm zp}$ & 24.9 & 27.6 \\
$t_{\rm exp}$ [sec] & 54 & 52 \\ 
Cadense [min] & 20 & 15 \\
SAA [deg] & 90-120 & 53-126 \\ \hline
\end{tabular}
\end{table}
% Euclid : $<33^{\circ}$ free insertion angle

\section{Microlensing event simulator} \label{sec:simulator}

\subsection{Event model} \label{subsec:model}
The basic structure of our microlensing simulator is almost the same as \citet{Ban2016, Ban2020}. We assumed the noise parameters from the \bes{} Galactic model version m1612 \citep{Marshall2006, Robin2003, Robin2012a, Robin2012b, Robin2017} and used the catalogues to draw sample source and lens properties except for replacing the lens mass to FFPs. The range of FFP mass is $10^{-6}M_{\odot}-10^{-2}M_{\odot}$ which roughly corresponds to 0.3 Earth-mass up to 10 Jupiter-mass FFPs. The target field is $(l,b)=(1.^{\circ}00,-1.^{\circ}75)$ where is expected to yield more events \citep{Penny2013, Ban2016}.

The full kinematics of the source, lens, and observers are considered in our simulator. Therefore, the parallax and xallarap are always included during the simulation process. In this research, we assume both source and FFP lens have no companions, so xallarap does not occur. Even though the parallax due to the observer's motion is simulated, we expect the parallax effect to be negligible for the FFP event duration.

\cite{Refsdal1966} discussed the two solutions of microlensing parallax and \cite{Gould2020} generalised the probability of two solutions. In the projected lens frame, there are two possible paths of the source that have the same minimum impact parameter (See also Figure 1 and 2 from \cite{Gould1994}). In our simulator, the two solutions of microlensing parallax are attributed to the net motion of the source, lens, and observer. Instead, either of the two paths of the source in the lens frame is randomly selected. We assume the probabilities of two paths are equal whilst the impact parameter value is quadratic from the centre of the lens.

\subsection{Telescopes} \label{subsec:telescope}
Table \ref{tab:survs} summarises the conditions and configurations of \euclid{} and \nancy{} in our microlensing event simulator. The \euclid{} orbit is assumed to take a maximum insertion angle ($\sim 33^{\circ}$) so that \euclid{} can fully take advantage of the solar aspect angle (SAA). As a result of these configurations, the cooperation period of \euclid{} and \nancy{} lasts for $\sim$40 days around equinoxes. \euclid{} has optical filters (VIS) and infrared filters (NISP) \citep{ESA2011}, and the infrared filter is much more suitable for FFP microlensing observation because it can cover the dimmer source regime and lower magnification events \citep{Bachelet2022}. \nancy{} is expected to use infrared filters for the exoplanet search via microlensing events \citep{Johnson2020, Spergel2013, Spergel2015}. Therefore, we assume the infrared filter for both \euclid{} and \nancy{}. Since our stellar catalogues from \bes{} Galactic model have the stellar magnitudes in Johnson-Cousin filters, we regard the $H$-band filter as a proxy of the \euclid{}'s and the \nancy{}'s infrared filters.

In the simulator, we manage the \euclid{} and \nancy{} kinematics by phasing the orbital motion from $-\pi$ to $\pi$ in the sky frame centring at the Lagrangian Point 2 (L2). When initialising the event properties, both \euclid{} and \nancy{} phases are randomly selected. The same phases between them are still acceptable because of the different orbital diameters. These selected phases are for the reference time ($t$=0) at which the impact parameter is also set up. We use \euclid{} as a reference observer because of the smaller SAA and brighter zero-point magnitude than those of \nancy{}. Thus, the \euclid{} sensitivity likely determines the simultaneous parallax detectability between them.

\subsection{Event criteria} \label{subsec:event}
We set the criteria of event alerting that the magnified flux results in the signal-to-noise ratio ($SN$) $>50$. The noise is parameterised in four ways: the background noise from unresolved stars ($m_{\rm stars}$), the blending noise from nearby stars ($m_{\rm blend}$), the detector noise by a point spread function (PSF) and sky brightness ($m_{\rm sky}$), and the photometric noise generalised by a square-root of source flux. Hence, the $SN$ equation becomes
\begin{equation}	\label{eq:sn}
\hspace{-0.1in}
S/N(t) = \frac{10^{0.2 m_{\rm zp}} \, t^{1/2}_{\rm exp} \, A(t) \, 10^{-0.4m_*}}{\sqrt{10^{-0.4 m_{\rm stars}} + 10^{-0.4 m_{\rm blend}} + \Omega_{\rm psf}10^{-0.4m_{\rm sky}} + A(t) \, 10^{-0.4m_*}}},
\end{equation}
where $A(t)$ is the amplitude for a given time. $m_{\rm zp}$, $t_{\rm exp}$, and $\Omega_{\rm psf}$ are the zero-point magnitude, exposure time, and the solid angle of the survey point spread function (PSF). All parameters except for the source baseline magnitude ($m_*$) are unique values depending on the telescope sensitivity (see Table \ref{tab:survs}). The unresolved background noise ($m_{\rm stars}$) and nearby bright star blending ($m_{\rm blend}$) are found by the same method as \citet{Ban2020} using \bes{} catalogue. The brief review of the calculation of $m_{\rm stars}$ and $m_{\rm blend}$ is as follows.

For a given source star ($j$), a boundary between resolved and unresolved magnitude is found. Supposing if the boundary magnitude is $m_{j_{lim}}$, then the total flux of stars fainter than the boundary magnitude ($i$) is $m_{\rm stars}$:
\begin{equation}	\label{eq:mstars}
B_{res,lim} = \ts\Omega_{\rm psf} \sum\limits_{m_i > m_{j_{lim}}}\frac{10^{-0.4m_i}}{\ts \Omega_{\rm cat,i}} = 10^{-0.4m_{\rm stars}},
\end{equation}
where $\Omega_{\rm cat}$ is the solid angle of the \bes{} data to adjust the population for different magnitudes. To find $j_{lim}$, we set the criteria that the given source star flux ($F_j$) over the noise from the total flux of unresolved stars ($\sqrt{B_{res}}$) is $>3$:
\begin{equation}	\label{eq:Bres}
\frac{F_j}{\sqrt{B_{res}}} = \frac{10^{0.2 m_{\rm zp}} \, t^{1/2}_{\rm exp} \, 10^{-0.4m_j}}{\sqrt{\ts\Omega_{\rm psf}\sum\limits_{m_i > m_j}\frac{10^{-0.4m_i}}{\ts \Omega_{\rm cat,i}}}} > 3.
\end{equation}

The blending noise by the nearby resolved star is also found using Eq.(\ref{eq:Bres}). For the blending, the total flux of the resolved stars within the PSF of the given star ($j$) becomes $m_{blend}$:
\begin{equation}	\label{eq:mblend}
B_{blend,j} = \ts\Omega_{\rm psf} \sum\limits_{nearby}\frac{10^{-0.4m_i}}{\ts \Omega_{\rm cat,i}} = 10^{-0.4m_{\rm blend}}.
\end{equation}
Note that the blending noise is a unique value for each source star in the catalogues per observer whilst the unresolved background noise is a unique value per observer.

\begin{figure}
\includegraphics[width=\textwidth]{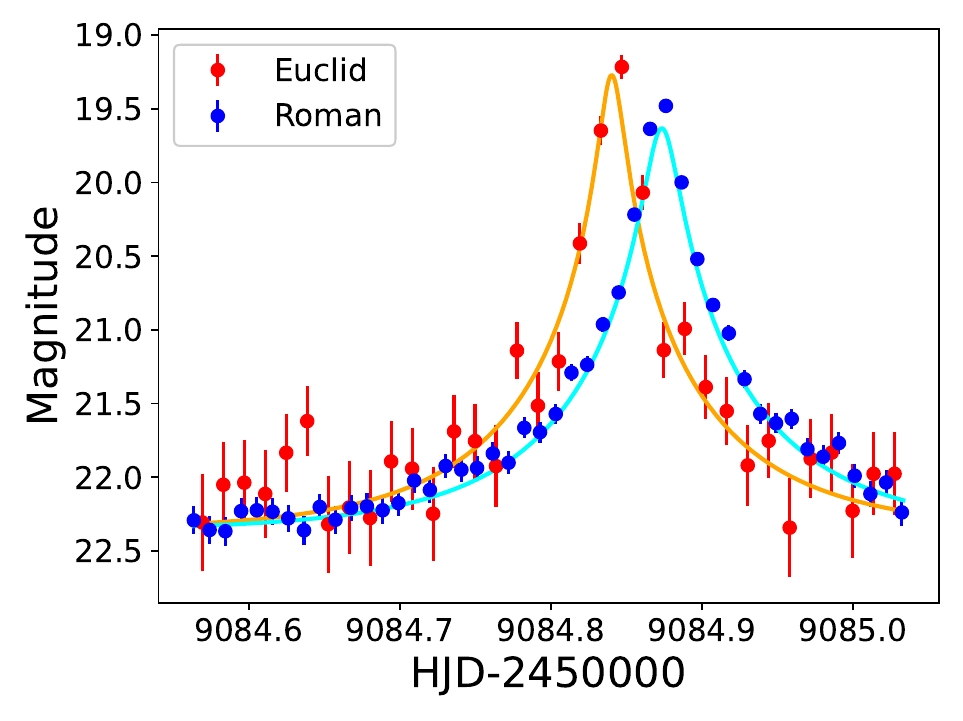} \\
\caption{An example of the photometric data observed by \euclid{} and \nancy{} in simultaneous parallax generated by our microlensing simulator.}
\label{fig:sample_lc}
\end{figure}

In addition to the alerting criteria of $SN>50$, we apply the least event duration limit of 1 hour. This means that the observed epochs that keep $SN>50$ must be lasting at least 1 hour in total. Note that the event duration is different from the Einstein timescale ($t_E = \theta_E / \mu_{rel}$). Once the event satisfies these criteria and is determined as an observable event, the weight of the event is calculated based on the event occurrence potential. We set a formula for the weight of the event as
\begin{equation}     \label{eq:wrate}
W_{SL} = u_{{\rm t},SL} D_L^2 \mu_{SL} \theta_{{\rm E},SL},
\end{equation}
where $SL$ stands for a given source-lens pair, $u_t$ is the threshold impact parameter derived from the threshold amplitude ($A_t$, see Table \ref{tab:survs}), $D_L$ is the lens distance, $\mu_{SL}$ is the source-lens relative proper motion, and $\theta_E$ is the Einstein radius \citep{Wambsganss1998}. As our previous research shows \citep{Ban2016}, $u_t$ varies between a point source and a finite source. For the point source event, there is a simple mutual relation between $A_t$ and $u_t$, 
\begin{equation}	\label{eq:au}
A(t) = \frac{u(t)^2+2}{u(t)\sqrt{u(t)^2+4}}.
\end{equation}
For the finite source event, Eq.(\ref{eq:au}) is no longer available. Our simulator uses the integration method over the apparent surface of the source for the finite source $A_t-u_t$ relation:
\begin{equation}	\label{eq:fs}
A(u(t)) = \frac{1}{\pi\rho^2}\int^{2\pi}_0\int^{\rho}_0 A[s(u(t),r,\phi)]r\dot dr d\phi,
\end{equation}
where $\rho$ is the angular radius of the source in units of $\theta_E$. It is also necessary to set up the criteria for parallax detectability in addition to event detectability. The parallax detectability is based on the differential light curve between two observers. the differential light curve is recognisable from any noises and uncertainties when either of the following conditions is satisfied:
\begin{equation}	\label{eq:dmax}
\left[\frac{dA(t)}{\sigma_{dA(t)}}\right]_{max} > 5,
\end{equation}
\begin{equation}	\label{eq:tmax}
\frac{dt_0}{\sigma_{dt_0}} > 5,
\end{equation}
where $dA(t)$ is the differential amplitude observed in two telescopes at a given time $t$, $dt_0$ is the time difference of the peak amplitude between two observers, and $\sigma$ is the uncertainty of those values. Figure \ref{fig:sample_lc} shows one of the sample parallax events generated from our microlensing simulator.

\begin{figure}
\includegraphics[width=\textwidth]{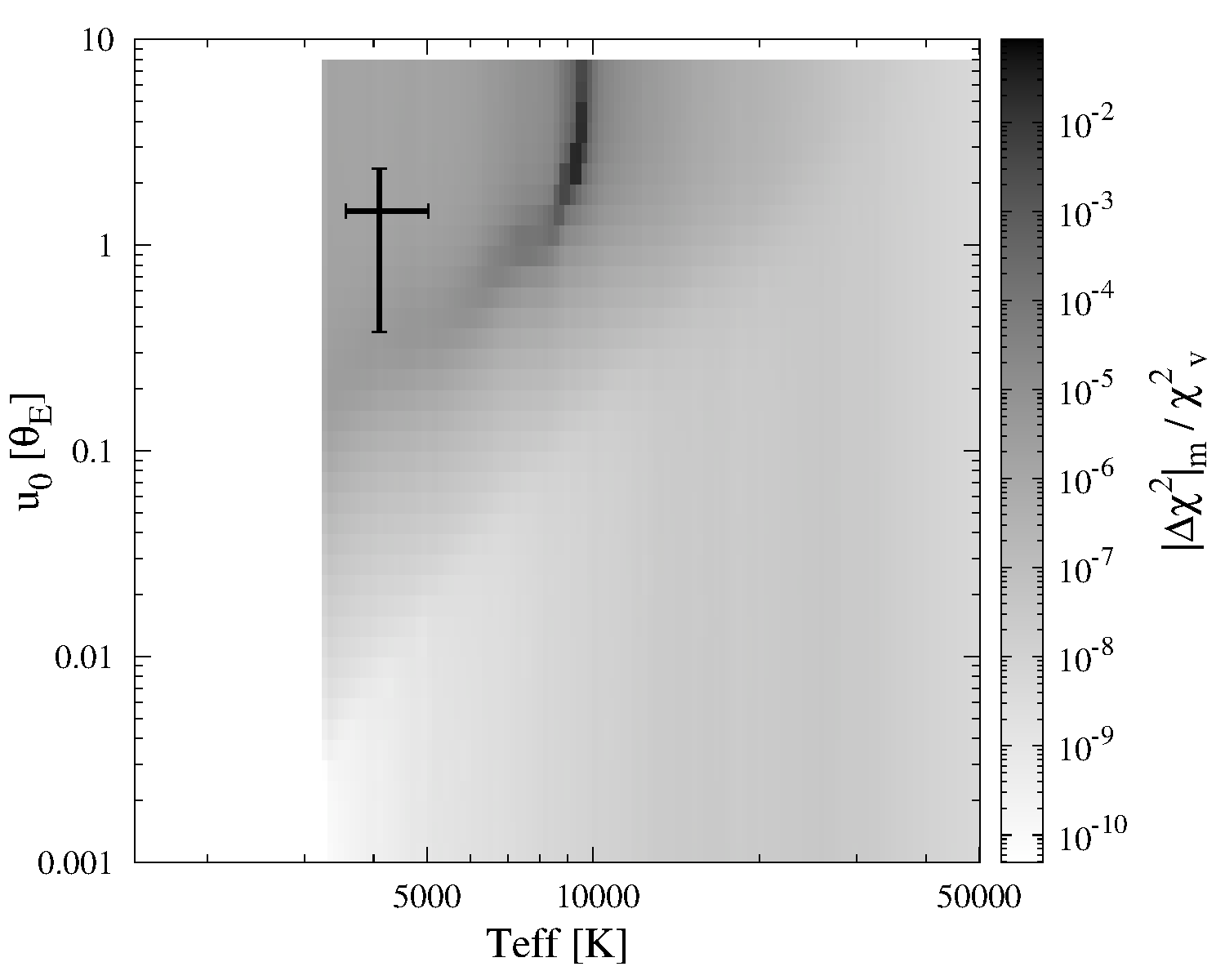} \\
\caption{The map of the ratio of the chi-square values for the different effective temperatures of a source and an impact parameter (see Eq.(\ref{eq:dchi2_ratio})). The limb-darkening coefficients are referred from \cite{Claret2012, Claret2017}. The finite source effects are simulated with the mean angular radii of the source derived from the \bes{} Galactic model catalogues assuming FFP lenses. The cross marker represents the weighted mean and $1\sigma$ distribution of $T_{eff}$ and $u_0$ of our sample events. Note that the uniform brightness was assumed for the sample drawing, and these sample events survived our detectability criteria (see \S\ref{subsec:event}).}
\label{fig:fse_ldc}
\end{figure}

For the FFP events, the finite source effect is never ignorable because of the smaller $\theta_E$ and larger $\rho$ than those of the stellar lens case \citep{Ban2016, Ban2020}. As Eq.(\ref{eq:fs}) shows, the amplitude with the finite source effect depends on the source surface brightness. The source surface brightness is usually not uniform, so we test the effect of limb darkening on the integration of the amplitude for the finite source effect. Figure \ref{fig:fse_ldc} shows a comparison of the $\chi^2$ values between the assumption of uniform brightness (UNI) and the limb-darkening coefficients (LDC). The colour bar indicates the ratio of $\chi^2$ derived as
\begin{equation}	\label{eq:dchi2_ratio}
\frac{|\Delta\chi^2|_m}{\chi^2_v} = \frac{|\chi^2_{m,LDC}-\chi^2_{m,UNI}|}{\chi^2_{v,UNI}},
\end{equation}
where the lower-case of $m$ stands for the discrepancy of data from a model whilst $v$ stands for the weighted sum of the squared variation. That is to say, $\chi^2_m$ presents how the theoretical light curve differs from the point-source-uniform-brightness source case, and $\chi^2_v$ presents the uncertainty of the observed photometry. As we can see in Figure \ref{fig:fse_ldc}, the light curve difference of the uniform brightness and limb-darkening cases is much smaller than the photometric uncertainty in the majority of events for FFP lenses. We also found that the $T_{eff}<3300$ [K] regime did not yield the finite source effect under the test conditions as Figure \ref{fig:fse_ldc}. Such a cool source likely be a low-mass dwarf having a small physical radius, so $\rho$ got small enough to be a point source. From this limb-darkening test, we finally determined that limb darkening is ignorable for our research of the accuracy of the microlensing parallax measurement. In Figure \ref{fig:fse_ldc}, we added the cross marker to briefly show the distribution of our sample events for the parallax measurement research that were assumed a uniform brightness case. The centre of the marker is at $(T_{eff},u_0)\sim(4127^{+1044}_{-497} K, 1.70^{+0.88}_{-1.31})$ with the $\chi^2$ ratio of $\sim 5.75\times 10^{-7}$.

\subsection{Microlensing parallax} \label{subsec:prlx_params}
The theoretical formula of the microlensing parallax ($\pi_E$) \citep{Refsdal1966, Gould1992} is
\begin{equation}	\label{eq:piE-theo}
\pi_E = \frac{AU}{\tilde{r_E}} = \sqrt{\kappa M_L D_{rel}},
\end{equation}
where
\begin{equation}	\label{eq:kappa}
\kappa = \frac{4G}{c^2},
\end{equation}
and
\begin{equation}	\label{eq:Drel}
\frac{1}{D_{rel}} = \frac{1}{D_L} - \frac{1}{D_S}.
\end{equation}
In the real observation, the value of microlensing parallax is estimated through the light curve fitting process, and the result is shown in 2D vectors in the projected lens frame. The vector components are perpendicular to each other and the line of sight. When the parallax effect is attributed mostly to the telescope separation, the vector components are set in the direction of the relative proper motion ($t$-component) and the direction of the minimum impact parameter ($u$-component). In this case, the parallax formula (Eq.(\ref{eq:piE-theo})) is rewritten as
\begin{equation}	\label{eq:piE-est}
\bm{\pi_E} = \frac{AU}{D_T}\left(\frac{\Delta t_0}{t_E} , \Delta u_0\right),
\end{equation}
where $D_T$ is the projected observer separation in units of AU perpendicular to the reference observer's line of sight, $t_0$ is the time when the amplitude peaks, $t_E$ is the Einstein timescale, and $u_0$ is the minimum impact parameter.

In this paper, we derive the theoretical $\pi_E$ with Eq.(\ref{eq:piE-theo}) using the values used for the event samples. The estimated $\pi_E$ is derived from the fitted parameter of the light curve and Eq.(\ref{eq:piE-est}). The estimation accuracy of the microlensing parallax is calculated by taking the discrepancies of the estimated $\pi_E$ from the theoretical $\pi_E$ as
\begin{equation}     \label{eq:discrepancy}
\Delta \pi_E [\%] = 100 \times \frac{|\bm{\pi_{E,est}}|-\pi_{E,th}}{\pi_{E,th}}.
\end{equation}

The weight of each sample event and any analytical constraints provide the fraction of events ($FoE$). It is the weighted ratio of the events that satisfy any analytical constraints over all events.
\begin{equation}     \label{eq:foe}
FoE(limit) = \frac{\sum_{SL,limit}^{N_{limit}} f(M_L)\cdot W_{SL}}{\sum_{sl}^n{f(M_l)\cdot W_{sl}}},
\end{equation}
where $f(M_L)$ is the population factor to adjust the sampling bias for different FFP mass functions. The detail of the population factor is described in the next section (\S\ref{sec:massfunc}). The ``limit'' indicates any analytical constraints such as the timescale, FFP mass, finite source effect, telescope phase difference, and so forth. The lower case of $s$, $l$, and $n$ in the denominator depends on the fraction type. If the fraction type is for all sample events, then $s=S$, $l=L$, $n=N$. If the fraction type is for all events (i.e. theoretically detectable by \euclid{} without considering the parallax and event duration criteria), then the lowercase iterations include both detectable and undetectable events under our parallax criteria. The fraction of events derived for any analytical constraints finally provides the event rate with the constraints as
\begin{equation}     \label{eq:eventrate}
\Gamma_{limit} = \Gamma_0 \times FoE(limit),
\end{equation}
where $\Gamma_0$ is the event rate of the theoretical solo observation. In this paper, the basic FFP population of the theoretical solo observation is that the 1 FFP per star and $\Gamma_0$ is derived with the same event detectability criteria (i.e. $S/N>50$ and event duration $>1$ hour) described in \S\ref{subsec:event}.

\section{FFP Mass functions} \label{sec:massfunc}
\begin{figure}
\includegraphics[width=\textwidth]{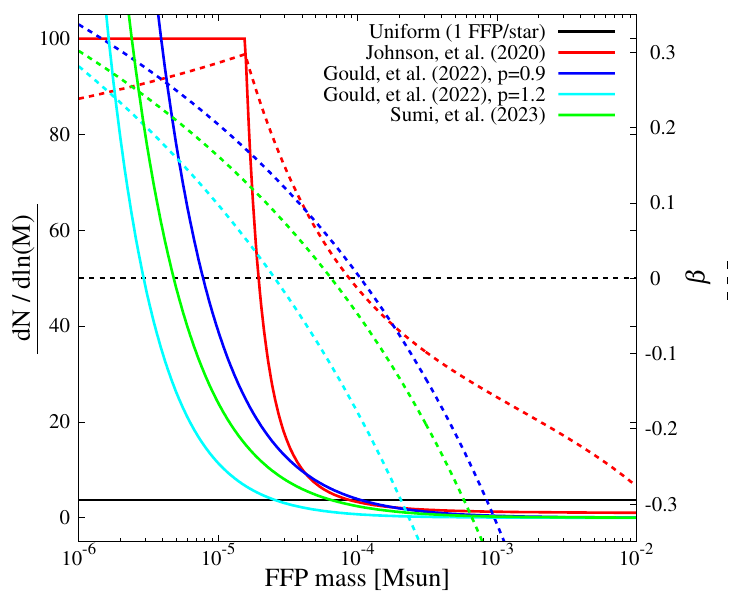} \\
\caption{Three different FFP mass functions in the solid curves and the slope of the population factor as a power law ($\beta$) in Eq.(\ref{eq:Pfactor}) in the dashed curves. The left y-axis corresponds to the mass function values, and the right y-axis corresponds to the $\beta$ value.}
\label{fig:functions}
\end{figure}

The variation of the FFP population is taken as the mass function slope ($\delta N/\delta ln M\propto M^{-\alpha}$). We use the population of 1 FFP per star as the standard mass function in our parallax simulation and label it as the ``Uniform'' mass function case. According to the stellar distribution of the \bes{} Galactic model catalogues, the Uniform function corresponds to $\delta N/\delta ln M \sim 3.7$ for our FFP mass range. In addition to the Uniform function, we consider four more mass functions as listed below.
%\begin{enumerate}[label=(\Alph*)]
\begin{itemize}
\item Uniform : $\delta N/\delta ln M \sim 3.7$
%\item \cite{Johnson2020} : $\delta N/\delta ln M = (0.00062/dex)\times10^{-0.73m}$ for $M_{FFP}/M_{\oplus}\leq 5.2$, 100 for $M_{FFP}/M_{\oplus}<5.2$
%\item \cite{Johnson2020}, Eq.(8) : $M_{FFP}/M_{\oplus}\geq5.2$ and $<5.2$ \\ \hspace*{0.5in}$\delta N/\delta ln M = (0.24/dex)\times(M_{FFP}/95M_{\oplus})^{-0.73}$\\ \hspace*{0.5in}$\delta N/\delta ln M = 100$
\item \cite{Johnson2020}, Eq.(8) : \\ \hspace*{1.0in}$\delta N/\delta ln M = (0.24/dex)\times(M_{FFP}/95M_{\oplus})^{-0.73}$ \hspace*{0.1in}for\hspace*{0.1in} $M_{FFP}/M_{\oplus}\geq5.2$ \\ \hspace*{1.0in}$\delta N/\delta ln M = 100$ \hspace*{0.1in}for\hspace*{0.1in} $M_{FFP}/M_{\oplus}<5.2$
\item \cite{Gould2022}, Eq.(10) with $p=0.9$ : \\ \hspace*{1.0in}$\delta N/\delta ln M = (0.4/dex)\times(M_{FFP}/38M_{\oplus})^{-0.9}$ \\
\item \cite{Gould2022}, Eq.(10) with $p=1.2$ : \\ \hspace*{1.0in}$\delta N/\delta ln M = (0.4/dex)\times(M_{FFP}/38M_{\oplus})^{-1.2}$ \\
\item \cite{Sumi2023}, Eq.(17) and Table 4 : \\ \hspace*{1.0in}$\delta N/\delta ln M \propto (1.85/dex)\times(M_{FFP}/8M_{\oplus})^{-1.14}$ \hspace*{0.1in}for\hspace*{0.1in} $M_{br} < M_{FFP}/M_{\odot} < 0.02$ \\ \hspace*{1.0in}$\delta N/\delta ln M \propto M_{FFP}^{-0.13}$ \hspace*{0.1in}for\hspace*{0.1in} $-7 < M_{FFP}/M_{\odot} < M_{br}$
\end{itemize}
%\end{enumerate}
where $m=log_{10}(M_{FFP}/M_{\odot})$, and the range is from -6 to -2 since our sample event FFP mass range is $10^{-6}M_{\odot}-10^{-2}M_{\odot}$. \cite{Johnson2020} estimated an FFP mass function with the FFP mass boundary at 5.2$M_{\oplus}$ based on the bound planet mass function by \cite{Cassan2012} and simulated the detectable event rate of FFP microlensing using \nancy{} configuration. \cite{Sumi2023} estimated the FFP mass function from the short events observed by \moa{}. As \cite{Sumi2023} mentioned about the total FFP mass per star, we need to adjust the event sampling bias between different mass functions for our research. Hence, we apply a population factor ($f(M_{FFP})$) against the Uniform function case assuming that the population factor forms a power law: 
\begin{equation}	\label{eq:Pfactor}
f(M_{FFP}) = \frac{N(M_{FFP})}{N(M_*)} = \left(\frac{M_{FFP}}{M_{\odot}}\right)^{-\beta}.
\end{equation}
For the Uniform function case, the population factor is 1 so that the slope of the power law ($\beta$) is 0 throughout the FFP mass. Figure \ref{fig:functions} visualises three mass functions and the slope of the power law of the population factor.

\section{Fitting the light curve of sample events using \mlm{}} \label{sec:fitting}
\mlm{} is one of the open-source tools for the microlensing light curve fitting \citep{Poleski2019}. We use this tool to find the estimated $\pi_E$ value for each sample event photometry. For all sample events, we consider (1) single-source-single-lens event, (2) finite-source effect, and (3) parallax by the observer's motion. The MCMC method is applied to find the best-fit parameters. Although \mlm{} has a function to include ephemeris of the space-based surveys, it works for a ground-based reference and space-based partner observers case. Therefore, we use another approach for the parallax by the motion of \euclid{} and \nancy{} in \mlm{}. Since we already expect that the light curve torsion due to the parallax is negligible for the FFP events because of the short event duration, we assume the motion of \euclid{} and \nancy{} during the observation of one event can be approximated to the shift of the L2 point; hence, the orbital motion of Earth. On the other hand, the impact parameter and the peak time of the light curve could largely differ between \euclid{} and \nancy{}, so we use the fitting steps as follows.
\begin{enumerate}
\item Fitting both \euclid{} and \nancy{} photometry together
\item Fitting each \euclid{} and \nancy{} photometry with a fixed value of $t_E$ and $\rho$ from the step 1
\item Checking reduced chi square ($\chi^2/d.o.f.$) for each \euclid{} and \nancy{}
\item Recording $t_0$, $u_0$, and $\pi_E$ for each \euclid{} and \nancy{} from the step 2 
\item Going back the step 1 with average $t_E$ and $\rho$ from the step 2
\item Repeating the step 1-5 until $\chi^2/d.o.f.<2$ for each \euclid{} and \nancy{}
\end{enumerate}
In the end, we gain a final record of $t_0$, $u_0$ and $\pi_E$ for each \euclid{} and \nancy{} individually and common $t_E$ and $\rho$ between them.

It is essential to derive both the microlensing parallax and the angular source radii from the light curve to estimate the lens properties. We assume the theoretical constraint of the recognisable finite source effect is $\rho/u_0<$1.0, and the sensitivity-based constraint is $\rho/u_0<$0.5 \citep{Johnson2020} for the reference observer \euclid{}. We consider both constraints of the angular source radii to calculate the accuracy of the microlensing parallax measurement from the fitted parameters.

The significance of the consideration of (2) finite-source effect and (3) parallax by the observer's motion are also confirmed by comparing $\chi^2/d.o.f.$ of with and without these considerations. As a result, $\chi^2/d.o.f.$ changes more significantly for the low-mass FFPs and small impact parameters. The comparison of the fitting results with and without the finite source consideration also takes part in the finite source constraints mentioned above. On the other hand, the observer's motion is ignorable as we expected and does not affect $\chi^2/d.o.f.$ and fitted parameters.

Although we optimised the parallax due to the observer's motion and did not include ephemeris in the light curve fitting process, we assume that the space-based surveys usually have a good record of the telescope coordinates in the sky. Our simulator yields the coordinates information for each detection epoch. The separation between \euclid{} and \nancy{} on the reference line ($D_T$) is derived from these coordinates.

\section{Results} \label{sec:results}

\subsection{Light curve fitting uncertainty}	\label{subsec:fitting_uncertainty}
The procedure of the light curve fitting shown in \S\ref{sec:fitting} yielded the best-fit parameters with uncertainties no greater than 20\%. We also calculated the discrepancy of the fitted parameters from the theoretical parameters used in the simulator. Those theoretical parameters were used as the initial parameters for the fitting process using \mlm{}. Table \ref{tab:fitting_discrepancy} shows the fitting parameter discrepancies and their likelihood among our samples for Earth-, Neptune-, and Jupiter-mass FFPs. The same concept as Eq.(\ref{eq:discrepancy}) were used to calculate the discrepancies for $t_E$, $\rho$, and $|u_0|$ for both \euclid{} and \nancy{}. The likelihood was the same as the fraction of the event which the fundamental formula is shown as Eq.(\ref{eq:foe}). The discrepancy of $t_0$ is removed from Table \ref{tab:fitting_discrepancy} because it reaches zero due to the \mlm{} fitting configurations, and the likelihood becomes $\sim$1. This is because we required all sample events to successfully observe the peak magnitude, so \mlm{} was given a true value of $t_0$, which also is directly readable from the input photometric data. Unlike the other event parameters such as $u_0$ and $t_E$, $t_0$ did not fluctuate so much during the fitting process as long as a given parameter and read data from the photometric data matches quite well.

Although we applied the theoretical parameters as the initial parameters of the light curve fitting, the MCMC approach yielded the best-fit parameters with some discrepancy from the theoretical values. For example, the Jupiter-mass FFP case likely yields a large discrepancy of the $\rho$ estimation. This is because $\rho$ of Jupiter-mass FFP generally becomes smaller than that of the low-mass FFPs so the discrepancy becomes relatively large. The comparison of the light curve fitting with and without finite source consideration does not show a large $\chi^2/d.o.f.$ difference for massive FFPs. Thus, there is a calibration limit of the small $\rho$ fitting, and the possible range of $\rho$ likely increases. Aside from the $\rho$ discrepancy, $t_E$ and $|u_0|$ are fitted well. The $t_E$ value of more than 80\% events and the $|u_0|$ value of around 90\% events are fitted within 100\% discrepancy. Here, the 100\% discrepancy corresponds to a factor of 2 uncertainty. Some sample events show $>$100\% discrepancy for the fitted parameters, and this is because of the number of epochs, magnification level, and lack of peak observation. Especially for the former two causes, those sample events passed our detectability criteria, but very barely.

%The likelihood shown in Table \ref{tab:fitting_discrepancy} is the result of the consistent approach of the light curve fitting listed in \S\ref{subsec:fitting}. The accuracy can be improved by customising the initialisation of the fitting and/or using several fitting tools to check the fitting plausibility for each event. Particularly, the $\rho$ value estimation for the massive FFPs can reach the same level of likelihood as that of the low-mass FFPs with the reasonable settings of parameter width for the MCMC approach.

\begin{table}
\centering
\caption{Likelihood of the fitted parameters of the \euclid{}-\nancy{} simultaneous microlensing observation for Earth-, Neptune-, and Jupiter-mass FFPs cases. The different discrepancy level ($<$10\%, $<$50\%, and $<$100\%) are extracted here. The discrepancy calculation follows the same concept as Eq.(\ref{eq:discrepancy}) for each parameter's theoretical and estimated values. The likelihood is derived as the fraction of events from the sample event weight over all samples. $E$ and $R$ notes for $|u_0|$ stand for \euclid{} and \nancy{}.}
\label{tab:fitting_discrepancy}
\begin{tabular}{|l|l|c|c|c|c|}
\hline
FFP mass & Discrepancy & $t_E$ & $\rho$ & $|u_{0,E}|$ & $|u_{0,R}|$ \\ \hline\hline
\multirow{3}{*}{Earth}	 & $<$10\%  & 0.25 & 0.10 & 0.15 & 0.18 \\
						 & $<$50\%  & 0.66 & 0.34 & 0.53 & 0.58 \\
						 & $<$100\% & 0.84 & 0.89 & 0.90 & 0.87 \\ \hline
\multirow{3}{*}{Neptune} & $<$10\%  & 0.35 & 0.12 & 0.23 & 0.29 \\
						 & $<$50\%  & 0.68 & 0.36 & 0.56 & 0.65 \\
						 & $<$100\% & 0.82 & 0.73 & 0.94 & 0.92 \\ \hline
\multirow{3}{*}{Jupiter} & $<$10\%  & 0.33 & 0.05 & 0.22 & 0.24 \\
						 & $<$50\%  & 0.69 & 0.22 & 0.57 & 0.63 \\
						 & $<$100\% & 0.85 & 0.48 & 0.93 & 0.92 \\ \hline
\end{tabular}
\end{table}

\subsection{Event rate and the accuracy of the microlensing parallax measurement} \label{subsec:eventrate_discre}
For a given FFP mass, the fraction of events for the simultaneous parallax observations does not change over the different FFP mass functions. We found that the probability of simultaneous parallax observation with finite source effect was $\sim$40\%, 16\%, and 4\% for Earth-mass, Neptune-mass, and Jupiter-mass FFPs under our criteria, respectively. Our parallax detectability criteria affect more on massive FFPs than low-mass FFPs since it focuses on the differential light curve. The criteria result in limiting the impact parameter of the Jupiter-mass FFP events. For instance, the differential magnitude of $u_{0,a}=0.5$ and $u_{0,b}=0.7$ is smaller than that of $u_{0,a}=0.1$ and $u_{0,b}=0.3$ even though the differences of the impact parameters are both 0.2. The latter case yields a larger magnitude difference and more possible to survive under our criteria. This tendency also relates to the occurrence of the finite source effect for Jupiter-mass FFPs. The Einstein ring of Jupiter-mass FFPs is generally large enough so it requires a small impact parameter to yield a finite source effect. On the other hand, Earth-mass FFPs yield a distinctive differential light curve more easily due to the small Einstein ring, and the finite source events are dominant \citep{Ban2020}.

\begin{figure*}
\centering
\includegraphics[width=\textwidth]{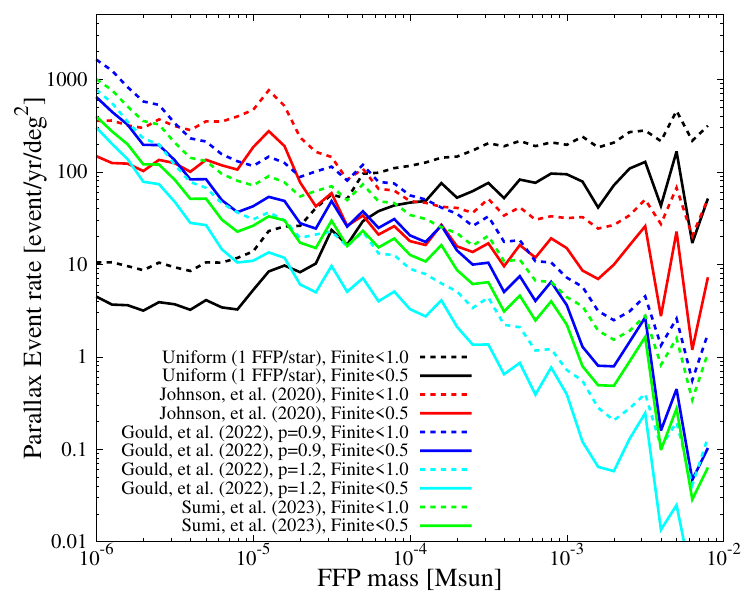} \\
\caption{Event rate [event/year/deg$^2$] for accurately measured microlensing parallax for different FFP mass functions. The FFP mass on the x-axis is binned by $dlog_{10}M_{FFP}\sim0.1$. The least acceptable discrepancy is assumed as $<100\%$, which corresponds to a factor of 2 uncertainty. The limitations of the finite source effect are $\rho/u_0<0.5$ in solid and $<1.0$ in dashed curves. Two solutions of the microlensing parallax are not considered in this plot and the $\delta u_0$ is taken from a typical subtraction formula.}
\label{fig:eventrate}
\end{figure*}

The event rate of accurately measured microlensing parallax varies with the FFP mass function because the population of a given FFP mass changes in the total FFP mass per star. Figure \ref{fig:eventrate} shows the event rate for three different FFP mass functions listed in \S\ref{sec:massfunc} along the FFP mass from $10^{-6}M_{\odot}$ to $10^{-2}M_{\odot}$. Two limits of finite source effect are plotted in the same diagram, and the discrepancy of $<100\%$ is assumed as the least acceptable level. Table \ref{tab:eventrate_ENJ} summarises the event rate of Earth-, Neptune-, and Jupiter-mass FFPs for three different mass functions. In this table, we assumed that the calculation of all events was done by a typical case (i.e. $\Delta u_0=u_{0,a}-u_{0,b}$ where $a$ and $b$ indicate separated observers \citep{Refsdal1966}) since it is usually difficult to distinguish between two solutions of the microlensing parallax from the light curve. In case the solutions are distinguished will be discussed in the later section \S\ref{sec:discussion}.

According to Figure \ref{fig:functions}, the FFP population density at $M_{FFP}\sim10^{-3.8\pm0.1}$ is almost agreed for all different FFP mass functions except for \citeauthor{Gould2022}'s function with $p=1.2$. However, the event rates for the mass in Figure \ref{fig:eventrate} do not agree with each other. This disagreement reflects the different population ratios of the mass to the total FFP mass per star. In other words, the optical depth for a given FFP mass lens and potential sources varies and determines the event rate. The \citeauthor{Gould2022}'s functions and \citeauthor{Sumi2023}'s function cross over the \citeauthor{Johnson2020}'s function at some FFP mass in Figure \ref{fig:functions}, and the FFP population densities roughly agree with each other. In this case, we can regard that the optical depth for the FFP mass lens and potential sources is almost equal between those mass functions.

\begin{table*}
\centering
\caption{Event rate [event/year/deg$^2$] for accurately measured microlensing parallax for different FFP mass functions. The numerical values are extracted for the conditions of Earth-, Neptune-, and Jupiter-mass FFPs, and $<$10\%, $<$50\%, and $<$100\% discrepancies. For each column of the mass function, the left sub-columns are for the limit of the finite source effect $\rho/u_0<$0.5, and the right sub-columns are for $<$1.0. Two solutions of the microlensing parallax are not considered in this plot and the $\delta u_0$ is taken from a typical subtraction formula.}
\label{tab:eventrate_ENJ}
\begin{tabular}{|l|l|cc|cc|cc|cc|cc|}
%\begin{tabular}{|l|l|>{\centering\arraybackslash}>{\centering\arraybackslash}p{0.055\textwidth}>{\centering\arraybackslash}p{0.055\textwidth}|>{\centering\arraybackslash}p{0.055\textwidth}>{\centering\arraybackslash}p{0.055\textwidth}|>{\centering\arraybackslash}p{0.055\textwidth}>{\centering\arraybackslash}p{0.055\textwidth}|>{\centering\arraybackslash}p{0.055\textwidth}>{\centering\arraybackslash}p{0.055\textwidth}|>{\centering\arraybackslash}p{0.055\textwidth}>{\centering\arraybackslash}p{0.055\textwidth}|}
\hline
\multirow{2}{*}{FFP mass} & \multirow{2}{*}{Discrepancy} & \multicolumn{2}{c|}{Uniform} & \multicolumn{2}{c|}{\citeauthor{Johnson2020}} & \multicolumn{2}{c|}{\citeauthor{Gould2022}, $p=0.9$} & \multicolumn{2}{c|}{\citeauthor{Gould2022}, $p=1.2$} & \multicolumn{2}{c|}{\citeauthor{Sumi2023}} \\ %\cdashline{3-12}
&& $<$0.5 & $<$1.0 & $<$0.5 & $<$1.0 & $<$0.5 & $<$1.0 & $<$0.5 & $<$1.0 & $<$0.5 & $<$1.0 \\ \hline\hline
\multirow{3}{*}{Earth}	 & $<$10\%  & 0.71 & 1.31 & 22.4 & 44.3 & 25.3 & 52.0 & 9.12 & 18.8 & 15.6 & 32.0 \\
						 & $<$50\%  & 2.68 & 6.24 & 87.0 & 208  & 101  & 242  & 36.7 & 87.2 & 62.4 & 149  \\
						 & $<$100\% & 3.74 & 9.53 & 126  & 325  & 147  & 379  & 53.2 & 137  & 90.7 & 233  \\ \hline
\multirow{3}{*}{Neptune} & $<$10\%  & 12.2 & 21.1 & 15.1 & 21.3 & 17.0 & 23.9 & 3.56 & 4.45 & 10.5 & 14.7 \\
						 & $<$50\%  & 25.9 & 77.9 & 28.8 & 85.1 & 32.2 & 94.4 & 6.08 & 17.7 & 19.8 & 58.1 \\
						 & $<$100\% & 30.2 & 95.5 & 32.2 & 103  & 36.0 & 114  & 6.72 & 21.3 & 22.1 & 70.1 \\ \hline
\multirow{3}{*}{Jupiter} & $<$10\%  & 24.4 & 69.5 & 4.86 & 13.7 & 1.36 & 3.15 & 0.15 & 0.30 & 0.84 & 1.94 \\
						 & $<$50\%  & 76.0 & 164  & 12.7 & 26.2 & 3.48 & 6.42 & 0.39 & 0.66 & 2.14 & 3.95 \\
						 & $<$100\% & 94.4 & 198  & 15.8 & 31.9 & 4.19 & 7.84 & 0.47 & 0.82 & 2.58 & 4.83 \\ \hline
\end{tabular}
\end{table*}

\section{Discussion}	\label{sec:discussion}

\subsection{Comparison with the other research} \label{subsec:comparison}
\cite{Bachelet2022} considered the probability of FFP lens mass and distance estimation assuming the mass function of 10 Earth-mass FFPs per Main-sequence star \citep{Mroz2019b} and 2 Jupiter-mass FFPs per Main-sequence stars \citep{Sumi2011}. They showed that 110 Earth-mass and 450 Jupiter-mass FFP microlensing events per year can be observed in parallax and $\sim$30 and $\sim$19 of them can determine the lens mass and distance within a factor of 2 uncertainty. Under the similar FFP population and total field of view (2.0 [deg$^2$]) as \cite{Bachelet2022}, our simulation indicates that 23 Earth-mass and 11 Jupiter-mass FFPs microlensing events per year are observable in parallax and $\sim$12 and $\sim$2 of them can determine the mass and distance within a factor of 2 uncertainty. Here, we assumed that the accuracy of the fitted parameters (see \S\ref{sec:fitting}) is improved, and 90\% of events resulted in all fitted parameters being within 100\% discrepancy for any FFP mass.

Our event rates are smaller than the values of \cite{Bachelet2022} because our observation condition is more limited, and detectability criteria are more strict than their configurations. Hence, to compare the results, we can consider the ratio of the detectable number of events that yields the event rate values above. Table 3 of \cite{Bachelet2022} shows the fraction of events (SN only):(SN+parallax):(SN+finite) $\sim$1:0.85:0.24 for Earth-mass and $\sim$1:0.92:0.04 for Jupiter-mass FFPs whilst our results are $\sim$1:0.69:0.58 and $\sim$1:0.20:0.18, respectively. For the Earth-mass FFP case, our criteria of parallax detectability are more strict than that of \cite{Bachelet2022} whilst a more finite source case is allowed. This is because our criteria take advantage of the extension of the threshold impact parameter with large $\rho$ (see \cite{Ban2016} for the details). As a total probability, \cite{Bachelet2022} shows $\sim$20\% of the Earth-mass FFP events and $\sim$4\% of the Jupiter-mass FFP events are observed in parallax, and the lens mass and distance can be determined in a factor of 2 uncertainty. Our corresponding results are $\sim$40\% of the Earth-mass and $\sim$4\% Jupiter-mass FFP events (see the discussion in \S\ref{subsec:eventrate_discre}). The larger probability of our result for the Earth-mass FFPs is due to the consideration of the extension of the threshold impact parameter which results in more detectable events satisfying the event duration limit of $>1$ hour. We also confirmed that the average expanded threshold impact parameter for Earth-mass FFPs is about twice as large as the point source case. Hence, it is plausible that our probability is doubled from \cite{Bachelet2022}. On the other hand, the Jupiter-mass FFP generally has less advantage of such extension of the threshold impact parameter, so the similar probability indicates that we reached a similar result to \cite{Bachelet2022} using the different approach of the event criteria.

\subsection{Two solutions of microlensing parallax} \label{subsec:delta_u0}
As \cite{Refsdal1966} explained, the two solutions of the microlensing parallax are attributed to the direction of the impact parameters. If the impact parameter vector is opposite between observers, then $\Delta u_0=u_{0,a}+u_{0,b}$ would be true; where $a$ and $b$ indicate separated observers. This equation makes the microlensing parallax larger than the typical subtraction of $\Delta u_0=u_{0,a}-u_{0,b}$. We checked all sample events in different FFP mass functions and found that the fraction of the addition case ($\Delta u_0(+)$) to the typical subtraction case ($\Delta u_0(-)$) is no greater than $1/10$ events for the FFP mass rage of $10^{-6}M_{\odot}-10^{-2}M_{\odot}$. The addition case is occurable for all FFP mass ranges in our research, and the low-mass FFPs are more possible. We also derived the event rate for accurately measured microlensing parallax if the solution is identified as either subtraction or addition cases for each sample event. If it is distinctive in either the subtraction or addition case, the event rate slightly increases from the case that all events are treated as the subtraction case (i.e. the results shown in \S\ref{sec:results}). However, the event rate becomes almost the same when the discrepancy of the microlensing parallax is accepted up to 100\%. Hence, it will be an advantage if either typical subtraction or addition calculation is distinctive, but it is not a big issue for FFPs under our criteria because of the low occurrence rate of an addition case.

\section{Conclusion}	\label{sec:conclusion}
We explored the influence of the different FFP mass functions on the event rate for accurately measured microlensing parallax and found that the event rate is related to the FFP population based on the extent of the finite source effect. For instance, the finite source event is dominant for the Earth-mass FFP events \citep{Ban2020}, and $\sim$40\% of them can measure the FFP mass and distance in a factor of 2 uncertainty. The probability becomes $\sim$16\% and $\sim$4\% for Neptune-mass and Jupiter-mass FFPs, respectively. However, the event rate does not simply rely on the number of FFPs for a given mass because the ratio of the FFP population over the total FFP mass per star determines the optical depth and event rate.

Aside from the influence of the different FFP mass functions on the event rate, the accuracy of the microlensing parallax measurement can be improved to some extent by improving the accuracy of the light curve fitting. Other than the technical issue of the fitting, the simple way of improvement is to use a higher cadence than we are expecting in our simulation or to have a multi-parallax of more than two telescopes. The ground-based surveys such as \moa{}, \ogle{}, and \kmt{} can also be partners for the parallax observation with any space-based telescopes \citep{Ban2016, Ban2020}. Moreover, there are some other telescopes planned to launch for L2 orbit such as \ariel{} and \plato{} \citep{Gardner2006, Rauer2014, Prusti2016, ESA2020}. Although the microlensing event is not a main target of these future surveys, they still have the potential to detect and alert microlensing events \citep{Mislis2016, Yee2017, Nikolaus2018}. By operating these space-based telescopes one after another, it would also be possible to overtake the parallax observation by the space-based surveys after the co-operation period of the \euclid{} and \nancy{}. It is necessary to gather the observation data to determine the FFP mass function, so a long-lasting mission as a whole is worthwhile.

%Rauer2014 PLATO
%ESA2020 Ariel
%Prusti2016 Gaia
%Gardner2006 JWST

\section{Acknowledgement}	\label{sec:ack}
Numerical computations were carried out on a PC cluster at the Center for Computational Astrophysics, National Astronomical Observatory of Japan. The work was supported by the Polish National Agency for Academic Exchange grant ``Polish Returns 2019'' to Rados\l{}aw Poleski.

\bibliography{ref}{}
\bibliographystyle{aasjournal}

\end{document}